\documentclass[showpacs,amsmath,amssymb,10pt,aps]{revtex4}
\usepackage{graphicx,color}
\usepackage{amsmath}
\usepackage{amssymb}
\usepackage{bbm, bm}
\usepackage{color}

\usepackage[T1]{fontenc}
\usepackage[cp1250]{inputenc}
\usepackage{amsfonts}
\usepackage{amssymb, amsbsy}
\usepackage{amsmath, bbm}
\usepackage{epstopdf}

\newtheorem{Theorem}{Theorem}

\begin{document}

\title{Quantum trajectories for environment in superposition of coherent states}

\author{Anita Magdalena D\k{a}browska}

\affiliation{Nicolaus Copernicus University in Toru\'{n},\\Collegium Medicum Bydgoszcz, ul. Jagiello\'{n}ska 15, 85-067 Bydgoszcz, Poland}

\begin{abstract}
We derive stochastic master equations for a quantum system interacting with a Bose field prepared in a superposition of continuous-mode coherent states. To determine a conditional evolution of the quantum system we use a collision model with an environment given as an infinite chain of not interacting between themselves qubits prepared initially in a entangled state being a discrete analogue of a superposition of coherent states of the Bose field. The elements of the environment chain interact with the quantum system in turn one by one and they are subsequently measured. We determine a conditional evolution of the quantum system for continuous in time observations of the output field as a limit of discrete recurrence equations. We consider the stochastic master equations for a counting as well as for a diffusive stochastic process.   
\end{abstract}

\maketitle


\section{Introduction}

Quantum filtering theory \cite{Bel89, Bel90, BB91, Car93, BP02, Bar06, GZ10, WM10} formulated within the framework of quantum stochastic It\^{o} calculus (QSC) \cite{HP84, Par92} gives the best state estimation of an open quantum system on the basis of a continuous in time measurement preformed on the Bose field interacting with the system. The filtering theory is formulated with the making use of input-output formalism \cite{GarCol85} wherein the input field is interpreted as the field before interaction with the system and the output field is interpreted as the field after this interaction. Information about the quantum system is gained in an indirect way by performing the measurements on the output field. In general, there are two types of the measurement considered in the filtering theory, namely, the photon counting and homodyne/heterodyne measurements which corresponds respectively to the counting and diffusion stochastic processes \cite{Bar06}. Evolution of an open quantum system conditioned on the results of the continuous in time measurement of the output field is given by the stochastic master equation called also in the literature the quantum filtering equation. The conditional state, depending on all past results of the measurement, creates quantum trajectory. By taking the average over all possible outcomes of the measurements we get from the {\it a posteriori} evolution the {\it a priori} evolution given by the master equation.  Clearly, the form of the filtering equation depends on the initial state of the environment and on the type of measurement performed on the output field. There exist many derivations of the filtering equations (see, for example, \cite{Bel89, Bel90, BB91, Car93, GG94, BP96, B02, GS04, WM10}). The rigorous derivations of the conditional evolution for the case when the Bose field is prepared in the Gaussian state one can find, for instance, in \cite{BGM04, GK10, DS11, DS12, N14, DG16}. The standard methods of determination of the filtering equation stop working when the Bose field is prepared in non-classical state. The initial temporal correlations in the Bose field makes then the evolution of open system non-Markovian. The system becomes entangled with the environment and its evolution is no longer given by one equation but by a set of equations. In this case to determine the conditional evolution of the system one can apply a cascaded approach \cite{GZ10} with an ancilla system being a source of non-classical signal. The methods of determination the filtering equation based on the idea of enlarging the Hilbert space of the compound system by the Hilbert space of ancilla were used for single photon state in \cite{ GJNC12a, GJN12b, GJN13, D18}, for a Fock state in \cite{GJN14,SZX16, BC17}, and for a superposition of coherent states in \cite{GJNC12a, GJN13}. Note, however that ancilla system serve here only as a convenient theoretical mathematical device allowing to solve the problem of determination of the conditional evolution. Unfortunately, by introducing such auxiliary system we loose some physical intuition and the interpretation of quantum trajectories become thereby more difficult.

In the paper we present derivation of the filtering equations for the environment prepared in a superposition of coherent states. Instead of the methods based on the concept of ancilla and QSC, we use quantum repeating interactions and measurements model  \cite{AP06, P08, PP09, P10}, known also in the physical literature as a collision model \cite{C17}.
We consider the environment modeled by an infinite chain of qubits which interact in turn one by one with a quantum system. After each interaction the measurement is preformed on the last qubit interacted with the system. The essential properties of our model are that each qubit interacts with the system only once and that the environment qubits do not interact between themselves. So in the paper we use the toy Fock space as an approximation of the symmetrical Fock space \cite{M93, At03, G04, GS04, AP06, BHJ09, P05}. The idea of obtaining the differential filtering equations from their difference versions were implemented for the Markovian case in \cite{B02, GS04, BHJ09, GCMC18}. As shown in \cite{DSC17, DSC18} it can be successfully applied also for the non-Markovian case.  

The paper is organized as follows. In Sec. II, we introduce a description of the environment and its interaction with the quantum system.  Sec. III is devoted to derivation of the conditional evolution of open system for the case when the environment is prepared in a coherent state. In Sec. IV the conditional evolution of open system for the bath in a superposition of coherent states is investigated. As an example, we present the {\it a priori} and the {\it a priori} dynamics of a single mode cavity in Sec. V. Our results are briefly summarized in Sec. VI.

\section{The unitary system and environment evolution}

Let us consider a quantum system $\mathcal{S}$ of the Hilbert space $\mathcal{H}_{\mathcal{S}}$ interacting with an environment consisting of a sequence of qubits. We assume that the environment qubits do not interact between themselves but they interact in a successive way with the system $\mathcal{S}$ each during the time interval of the length $\tau$. At a given moment $\mathcal{S}$ interacts with only one of the environment qubits. The Hilbert space of the environment is
\begin{equation}
\mathcal{H}_{\mathcal{E}}=\bigotimes_{k=0}^{+\infty}\mathcal{H}_{\mathcal{E},k},
\end{equation}
where $\mathcal{H}_{\mathcal{E},k}$ stands for the Hilbert space of the $k$-th qubit interacting with $\mathcal{S}$ in the time interval $[k\tau, (k+1)\tau)$. We start from a discrete in time model of repeated interactions (collisions) to show finally its limit with time treated as a continuous variable. We will treat $\tau$ as a small time and work to linear order in $\tau$ (we neglect all higher order terms in $\tau$).     

We assume that the unitary evolution of the compound $\mathcal{E}+\mathcal{S}$ system is governed by \cite{AP06,PP09}
\begin{equation}
U_{j}=\mathbb{V}_{j-1}\mathbb{V}_{j-2}\ldots \mathbb{V}_{0}\;\; \mathrm{for}\;\; j\geq 1, \;\;\;\;\;U_{0}=\mathbbm{1},
\end{equation}
where $\mathbb{V}_{k}$ is the unitary operator acting non-trivially only in the Hilbert space $\mathcal{H}_{\mathcal{E},k}\otimes \mathcal{H}_{S}$, that is,
\begin{equation}\label{intermat}
\mathbb{V}_{k}= \bigotimes_{i=0}^{k-1}\mathbbm{1}_{i} \otimes {V}_{k},
\end{equation}
and
\begin{equation}
{V}_{k} =  \exp\left(-i\tau H_{k}\right),
\end{equation}
with
\begin{eqnarray}\label{hamint}
H_{k} = \bigotimes_{i=k}^{+\infty}\mathbbm{1}_{i}\otimes H_{\mathcal{S}}+\frac{i}{\sqrt{\tau}}\left(\sigma_{k}^{+}\otimes \bigotimes_{i=k+1}^{+\infty}\mathbbm{1}_{i}\otimes L-\sigma_{k}^{-}\otimes \bigotimes_{i=k+1}^{+\infty}\mathbbm{1}_{i}\otimes L^{\dagger}\right),
\end{eqnarray}
where $H_\mathcal{S}$ is the Hamiltonian of $\mathcal{S}$, $L$ is a bounded operator of $\mathcal{S}$, and $\sigma_{k}^{+}=|1\rangle_{k}\langle 0|$, $\sigma_{k}^{-}=|0\rangle_{k}\langle 1|$, where by $|0\rangle_{k}$ and $|1\rangle_{k}$ we denoted respectively the ground and excited states of the $k$-th qubit. The Hamiltonian $H_{k}$ is written in the interaction picture eliminating the free evolution of the bath. A detailed discussion on the physical assumptions leading to (\ref{hamint}) one can find, for instance, in \cite{C17,GCMC18,FTVRS18}. For simplicity, we set the Planck constant $\hbar=1$. Note that $U_{j}$ describes the $j$-th first interactions and it has trivial action on $\bigotimes_{k=j}^{+\infty}\mathcal{H}_{\mathcal{E},k}$. 

Let us define in $\mathcal{H}_{\mathcal{E},k}$ the vector $|\alpha_{k}\rangle_{k}$ by the formula \cite{GCMC18}
\begin{equation}
|\alpha_{k}\rangle_{k}=e^{\sqrt{\tau}\left(\alpha_{k}\sigma_{k}^{+}-\alpha_{k}^{\ast}\sigma_{k}^{-}\right)}|0\rangle_{k},
\end{equation}
where $\alpha_{k}\in\mathbb{C}$. One can check that 
\begin{equation}
|\alpha_{k}\rangle_{k}=\left(1-\frac{|\alpha_{k}|^2}{2}\tau\right)|0\rangle_{k}+\alpha_{k}\sqrt{\tau}|1\rangle_{k}+O(\tau^{3/2}) 
\end{equation}
and 
\begin{equation}
\langle \alpha_{k}|\sigma_{k}^{-}|\alpha_{k}\rangle=\sqrt{\tau}\alpha_{k}+O(\tau^{3/2}),\;\;
\langle \alpha_{k}|\sigma_{k}^{+}\sigma_{k}^{-}| \alpha_{k}\rangle=\tau|\alpha_{k}|^2+O(\tau^2).
\end{equation}
The coherent state in $\mathcal{H}_{\mathcal{E}}$ we define as 
\begin{equation}
|\alpha\rangle=
\displaystyle{\bigotimes_{k=0}^{+\infty}}|\alpha_{k}\rangle_{k}
\end{equation}
with the condition $\displaystyle{\sum_{k=0}^{+\infty}}|\alpha_{k}|^2\tau<\infty$.

Note that the vector state $|\alpha\rangle$ is a discrete analogue of coherent state defined in the symmetric Fock space considered in QSC. We will show that it allows in the continuous time limit to reproduce all results for the coherent state received within QSC.

\section{Quantum trajectories for the coherent state}

In this section we consider the case when the composed $\mathcal{E}+\mathcal{S}$ system is prepared initially in the pure product state
\begin{equation}\label{ini1}
|\alpha\rangle\otimes|\psi\rangle, 
\end{equation}
where $|\alpha\rangle$ is the coherent state of the environment.

\subsection{Photon counting}

 We assume that after each interaction the measurement is performed on the last element of the environment chain just after its interaction with $\mathcal{S}$.  A goal of this subsection is providing a description of the state of $\mathcal{S}$ conditioned on the results of the measurements of the observables 
\begin{equation}\label{obser}
\sigma_{k}^{-}\sigma_{k}^{+}=|1\rangle_{k}\langle 1|,\;\;\;k=0,1,2,\ldots .
\end{equation} 

\begin{Theorem}\label{TH-1}
The conditional state of $\mathcal{S}$ and the part of the environment which has not interacted with $\mathcal{S}$ up to $j\tau$ for the initial state (\ref{ini1}) and the measurement of (\ref{obser}) at the moment $j\tau$ is given by
\begin{equation}\label{cond}
|\tilde{\Psi}_{j}\rangle = \frac{|\Psi_{j}\rangle}{\sqrt{\langle\Psi_{j}|\Psi_{j}\rangle}},
\end{equation}
where
\begin{equation}\label{cond1}
|\Psi_{j}\rangle=\bigotimes_{k=j}^{+\infty}|\alpha_{k}\rangle_{k}\otimes|\psi_{j}\rangle
\end{equation}
and the conditional vector $|\psi_{j}\rangle$ from $\mathcal{H}_{S}$ satisfies the recurrence formula
\begin{equation}\label{recur}
|\psi_{j+1}\rangle=M_{\eta_{j+1}}^{j}|\psi_{j}\rangle,
\end{equation}
where $\eta_{j+1}$ stands for a random variable describing the $(j+1)$-th output of (\ref{obser}), and  $M_{\eta_{j+1}}^{j}$ has the form
\begin{eqnarray}
M_{0}^{j}&=&\mathbbm{1}_{S}-\left(iH_{\mathcal{S}}+\frac{1}{2}L^{\dagger}L+L^{\dagger}\alpha_{j}+\frac{|\alpha_{j}|^2}{2}\right)\tau+O(\tau^2),\\
M_{1}^{j}&=&\left(L+\alpha_{j}\right)\sqrt{\tau}+O(\tau^{3/2}). 
\end{eqnarray}
Initially $|\psi_{j=0}\rangle=|\psi\rangle$ such that $|\tilde{\Psi}_{j=0}\rangle=|\alpha\rangle\otimes|\psi\rangle$. 
\end{Theorem}

\noindent It is clear that  $|\tilde{\Psi}_{j}\rangle$ is the product state vector belonging to the Hilbert space $\displaystyle{\bigotimes_{k=j}^{+\infty}}\mathcal{H}_{\mathcal{E},k}\otimes \mathcal{H}_{S}$. Note also that the conditional vector $|\psi_{j}\rangle$ depends on all results of the measurements performed on the bath qubits up to time $j\tau$.

{\it Proof.} We prove the above theorem by an induction technique. So we start from the assumption that (\ref{cond1}) holds and then check that 
	\begin{eqnarray}\label{action}
V_{j}|\Psi_{j}\rangle&=&|0\rangle_{j}\otimes\bigotimes_{k=j+1}^{+\infty}|\alpha_{k}\rangle_{k}\otimes
\left(\mathbbm{1}_{S}-\left(iH_{\mathcal{S}}+\frac{1}{2}L^{\dagger}L+L^{\dagger}\alpha_{j}
+\frac{|\alpha_{j}|^2}{2}\right)\tau+O(\tau^2)\right)|\psi_{j}\rangle\nonumber\\
&&+|1\rangle_{j}\otimes\bigotimes_{k=j+1}^{+\infty}|\alpha_{k}\rangle_{k}\otimes \left[\left(L+\alpha_{j}\right)\sqrt{\tau}+O(\tau^{3/2})\right]|\psi_{j}\rangle.
\end{eqnarray}
Now using the fact that the conditional vector $|\Psi_{j+1}\rangle$ from the Hilbert space $\displaystyle{\bigotimes_{k=j+1}^{+\infty}}\mathcal{H}_{\mathcal{E},k}\otimes \mathcal{H}_{S}$
is defined by
\begin{equation}
\left(\Pi_{\eta_{j+1}}^{j}\otimes\displaystyle{\bigotimes_{k=j+1}^{+\infty}}\mathbbm{1}_{k}\otimes \mathbbm{1}_{S}\right)V_{j}|\Psi_j\rangle=|\eta_{j+1}\rangle_{j}\otimes|\Psi_{j+1}\rangle,
\end{equation}
where
\begin{equation}
\Pi_{0}^{j}=|0\rangle_{j}\langle 0|,\;\;\;\;
\Pi_{1}^{j}=|1\rangle_{j}\langle 1|,
\end{equation}
we readily find that $|\Psi_{j+1}\rangle$ has the form 
\begin{equation}
|\Psi_{j+1}\rangle=\bigotimes_{k=j+1}^{+\infty}|\alpha_{k}\rangle_{k}\otimes|\psi_{j+1}\rangle
\end{equation}
with $|\psi_{j+1}\rangle$ given by (\ref{recur}), which ends the proof.

\subsection{Homodyne detection}

Now we describe the evolution conditioned on the results of the measurements of the observables 
\begin{equation}\label{obs2}
\sigma_{k}^{x}=
\sigma_{k}^{+}+\sigma_{k}^{-}=|+\rangle_{k}\langle+|-|-\rangle_{k}\langle-| ,\;\;\;k=0,1,2,\ldots,
\end{equation}
where 
\begin{eqnarray}\label{xbase}
|+\rangle_{k} &=&\frac{1}{\sqrt{2}}\left(|0\rangle_{k}+|1\rangle_{k}\right),\\
|-\rangle_{k} &=&\frac{1}{\sqrt{2}}\left(|0\rangle_{k}-|1\rangle_{k}\right),
\end{eqnarray}
are vectors from the Hilbert space $\mathcal{H}_{\mathcal{E},k}$.

\begin{Theorem}\label{TH-2} The conditional state of $\mathcal{S}$ and the part of the environment which has not interacted with $\mathcal{S}$ up to $j\tau$ for the initial state (\ref{ini1}) and the measurement of (\ref{obs2}) at the moment $j\tau$ is given by
\begin{equation}\label{condb1}
|\tilde{\Psi}_{j}\rangle = \frac{|\Psi_{j}\rangle}{\sqrt{\langle\Psi_{j}|\Psi_{j}\rangle}},
\end{equation}
where
\begin{equation}\label{condb2}
|\Psi_{j}\rangle=\bigotimes_{k=j}^{+\infty}|\alpha_{k}\rangle_{k}\otimes|\psi_{j}\rangle
\end{equation}
and the conditional vector $|\psi_{j}\rangle$ from $\mathcal{H}_{S}$ satisfies the recurrence formula
\begin{equation}\label{recurb}
|\psi_{j+1}\rangle=R_{\zeta_{j+1}}^{j}|\psi_{j}\rangle,
\end{equation}
where $\zeta_{j+1}=\pm 1$ stands for a random variable describing $(j+1)$-th output of (\ref{obs2}), and  
\begin{eqnarray}\label{Mq}
R_{\zeta_{j+1}}^{j}=\frac{1}{\sqrt{2}}\left[\mathbbm{1}_{S}-\left(iH_{\mathcal{S}}+\frac{1}{2}L^{\dagger}L+L^{\dagger}\alpha_{j}+\frac{|\alpha_{j}|^2}{2}\right)\tau
+(L+\alpha_{j})\zeta_{j+1}\sqrt{\tau}+O(\tau^{3/2})\right].
\end{eqnarray}
Initially $|\psi_{j=0}\rangle=|\psi\rangle$ such that $|\tilde{\Psi}_{j=0}\rangle=|\alpha\rangle\otimes|\psi\rangle$. 
\end{Theorem}	

{\it Proof}. Assuming that (\ref{condb2}) holds we get  
\begin{eqnarray}\label{action2}
V_{j}|\Psi_{j}\rangle&=&\frac{1}{\sqrt{2}}|+\rangle_{j}\otimes\bigotimes_{k=j+1}^{+\infty}|\alpha_{k}\rangle_{k}\otimes
\left\{\mathbbm{1}_{S}-\left(iH_{\mathcal{S}}+\frac{1}{2}L^{\dagger}L+L^{\dagger}\alpha_{j}
+\frac{|\alpha_{j}|^2}{2}\right)\tau\right.\nonumber\\&&\left.+\left(L+\alpha_{j}\right)\sqrt{\tau}+O(\tau^{3/2})\right\}|\psi_{j}\rangle\nonumber\\
&&+\frac{1}{\sqrt{2}}|-\rangle_{j}\otimes\bigotimes_{k=j+1}^{+\infty}|\alpha_{k}\rangle_{k}\otimes \left\{\mathbbm{1}_{S}-\left(iH_{\mathcal{S}}+\frac{1}{2}L^{\dagger}L+L^{\dagger}\alpha_{j}
+\frac{|\alpha_{j}|^2}{2}\right)\tau\right.\nonumber\\&&\left.-\left(L+\alpha_{j}\right)\sqrt{\tau}+O(\tau^{3/2})\right\} |\psi_{j}\rangle.
\end{eqnarray}
The conditional vector $|\Psi_{j+1}\rangle$ from the Hilbert space $\displaystyle{\bigotimes_{k=j+1}^{+\infty}}\mathcal{H}_{\mathcal{E},k}\otimes \mathcal{H}_{S}$ satisfies for the measurement of (\ref{obs2})  the equation 
\begin{equation}
\left(\Pi_{\zeta_{j+1}}^{j}\otimes\displaystyle{\bigotimes_{k=j+1}^{+\infty}}\mathbbm{1}_{k}\otimes \mathbbm{1}_{S}\right)V_{j}|\Psi_j\rangle=| \zeta_{j+1}\rangle_{j}\otimes|\Psi_{j+1}\rangle,
\end{equation}
where $\zeta_{j+1}$ has to possible values $\pm 1$, and
\begin{equation}
\Pi_{+1}^{j}=|+\rangle_{j}\langle +|,\;\;\;\; \Pi_{-1}^{j}=|-\rangle_{j}\langle -|.
\end{equation}
It is seen that $|\Psi_{j+1}\rangle$ has the form of (\ref{condb2}) and the vector $|\psi_{j}\rangle$ from $\mathcal{H}_{\mathcal{S}}$ satisfies the recurrence equation (\ref{Mq}).

\section{Quantum trajectories for a superposition of coherent states}

Let us assume that the initial state of the compound $\mathcal{E}+\mathcal{S}$ system is given by
\begin{equation}\label{ini2}
\left(c_{\alpha}|\alpha\rangle+c_{\beta}|\beta\rangle\right)\otimes|\psi\rangle,
\end{equation}
where
$|\alpha\rangle$ and $|\beta\rangle$ are coherent states of $\mathcal{H}_{\mathcal{E}}$, and 
\begin{equation}
|c_{\alpha}|^2+c_{\alpha}c_{\beta}^{\ast}\langle\beta|\alpha\rangle+c_{\alpha}^{\ast}c_{\beta}\langle\alpha|\beta\rangle+|c_{\beta}|^2=1. 
\end{equation}
Note that in this case the bath qubits are prepared in the entangled state. 

\subsection{Photon counting}

\begin{Theorem}\label{TH-3} The conditional state of $\mathcal{S}$ and the part of the environment which has not interacted with $\mathcal{S}$ up to $j\tau$ for the initial state (\ref{ini2}) and the measurement of (\ref{obser}) at the moment $j\tau$ is given by
\begin{equation}\label{cond2}
|\tilde{\Psi}_{j}\rangle = \frac{|\Psi_{j}\rangle}{\sqrt{\langle\Psi_{j}|\Psi_{j}\rangle}},
\end{equation}
where
\begin{equation}\label{cond4}
|\Psi_{j}\rangle=c_{\alpha}\bigotimes_{k=j}^{+\infty}|\alpha_{k}\rangle_{k}\otimes|\psi_{j}\rangle+
c_{\beta}\bigotimes_{k=j}^{+\infty}|\beta_{k}\rangle_{k}\otimes|\varphi_{j}\rangle.
\end{equation}
The conditional vectors $|\psi_{j}\rangle$, $|\varphi_{j}\rangle$ from $\mathcal{H}_{\mathcal{S}}$ in (\ref{cond4}) are given by the recurrence formulas
\begin{equation}\label{rec1}
|\psi_{j+1}\rangle=M_{\eta_{j+1}}^{\alpha_{j}}|\psi_{j}\rangle,
\end{equation}
\begin{equation}\label{rec2}
|\varphi_{j+1}\rangle=M_{\eta_{j+1}}^{\beta_{j}}|\varphi_{j}\rangle,
\end{equation}
where $\eta_{j+1}=0,1$ stands for a random variable describing the $(j+1)$-th output of (\ref{obser}), and
\begin{equation}
M_{0}^{\alpha_{j}}=\mathbbm{1}_{S}-\left(iH_{\mathcal{S}}+\frac{1}{2}L^{\dagger}L+L^{\dagger}\alpha_{j}+\frac{|\alpha_{j}|^2}{2}\right)\tau +O(\tau^2),
\end{equation}
\begin{equation}
M_{0}^{\beta_{j}}=\mathbbm{1}_{S}-\left(iH_{\mathcal{S}}+\frac{1}{2}L^{\dagger}L+L^{\dagger}\beta_{j}+\frac{|\beta_{j}|^2}{2}\right)\tau +O(\tau^2),
\end{equation}
\begin{equation}
M_{1}^{\alpha_{j}}=\left(L+\alpha_{j}\right)\sqrt{\tau}+O(\tau^{3/2}),
\end{equation}
\begin{equation}
M_{1}^{\beta_{j}}=\left(L+\beta_{j}\right)\sqrt{\tau}+O(\tau^{3/2}),
\end{equation}
and initially we have $|\psi_{0}\rangle=|\varphi_{0}\rangle=|\psi\rangle$. 
\end{Theorem}

\noindent {\it Proof}. The proof is straightforward. We simply refer to the results of the previous Section and the linearity of the evolution of the total system. 

\vspace{0.5cm}

Let us notice that the form of $|\Psi_{j}\rangle$ indicates that the system $\mathcal{S}$ becomes entangled with this part of the environment which has not interacted with $\mathcal{S}$ yet. Taking the partial trace of the operator $|{\Psi}_{j}\rangle\langle{\Psi}_{j}|$ over $\mathcal{S}$ we get the unnnormalized state of the environment of the form
\begin{eqnarray}
\rho^{field}_{j}=\lefteqn{|c_{\alpha}|^2\bigotimes_{k=j}^{+\infty}|\alpha_{k}\rangle_{k}\langle\alpha_{k}|\langle\psi_{j}|\psi_{j}\rangle
	+c_{\alpha}c_{\beta}^{\ast}\bigotimes_{k=j}^{+\infty}|\alpha_{k}\rangle_{k}\langle\beta_{k}|\langle
	\varphi_{j}|\psi_{j}\rangle}\nonumber\\
&&\!+c_{\alpha}^{\ast}c_{\beta} \bigotimes_{k=j}^{+\infty}|\beta_{k}\rangle_{k}\langle\alpha_{k}|\langle
\psi_{j}|\varphi_{j}\rangle
\!+\!|c_{\beta}|^2|\bigotimes_{k=j}^{+\infty}|\beta_{k}\rangle_{k}
\langle\beta_{k}|\langle\varphi_{j}|\varphi_{j}\rangle. 
\end{eqnarray}
The operator $\rho^{field}_{j}$ describes the conditional state of this part of the environment which has not interacted with $\mathcal{S}$ yet. It depends on all results of the measurements performed on the bath qubits up to $j\tau$. Therefore, we can say that the results of the measurements changes our knowledge about the state of the future part of the environment. 

In order to obtain the conditional state of $\mathcal{S}$ one has to take the partial trace of $|\tilde{\Psi}_{j}\rangle\langle\tilde{\Psi}_{j}|$ over the environment. One can check that the {\it a posteriori} state of $\mathcal{S}$ at the time $j\tau$ has the form
\begin{equation}
\tilde{\rho}_{j}
=\frac{\rho_{j}}{\mathrm{Tr}\rho_{j}},
\end{equation}
where
\begin{eqnarray}
\rho_{j}&=&|c_{\alpha}|^2|\psi_{j}\rangle\langle\psi_{j}|+c_{\alpha}c_{\beta}^{\ast}\prod_{k=j}^{+\infty}\langle \beta_{k}|\alpha_{k}\rangle|\psi_{j}\rangle\langle\varphi_{j}|+c_{\alpha}^{\ast}c_{\beta}\prod_{k=j}^{+\infty}\langle \alpha_{k}|\beta_{k}\rangle|\varphi_{j}\rangle\langle\psi_{j}|+|c_{\beta}|^2|\varphi_{j}\rangle\langle\varphi_{j}|
\end{eqnarray}
and $\mathrm{Tr}\rho_{j}$ is the probability of a particular trajectory. 
 
To derive the set of recurrence equations describing the stochastic evolution of $\mathcal{S}$ it is convenient to write down the conditional state of $\mathcal{S}$ at $j\tau$ in the form
\begin{equation}\label{Filter}
\tilde{\rho}_{j}=
|c_{\alpha}|^2\tilde{\rho}_{j}^{\alpha\alpha}+c_{\alpha}c_{\beta}^{\ast}\tilde{\rho}_{j}^{\alpha\beta}
+c_{\alpha}^{\ast}c_{\beta}\tilde{\rho}_{j}^{\beta\alpha}
+|c_{\beta}|^2\tilde{\rho}_{j}^{\beta\beta},
\end{equation}
where
\begin{equation}\label{oper1}
\tilde{\rho}_{j}^{\alpha\alpha}=\frac{1}{\mathrm{Tr}\rho_{j}}|\psi_{j}\rangle\langle\psi_{j}|,
\end{equation}
\begin{equation}\label{oper2}
\tilde{\rho}_{j}^{\alpha\beta}=\frac{\displaystyle{\prod_{k=j}^{+\infty}} {}_{k}\langle \beta_{k}|\alpha_{k}\rangle_{k}}{\mathrm{Tr}\rho_{j}}|\psi_{j}\rangle\langle\varphi_{j}|,
\end{equation}
\begin{equation}\label{oper3}
\tilde{\rho}_{j}^{\beta\alpha}=\frac{\displaystyle{\prod_{k=j}^{+\infty}} {}_{k}\langle \alpha_{k}|\beta_{k}\rangle_{k}}{\mathrm{Tr}\rho_{j}}|\varphi_{j}\rangle\langle\psi_{j}|,
\end{equation}
\begin{equation}\label{oper4}
\tilde{\rho}_{j}^{\beta\beta}=\frac{1}{\mathrm{Tr}\rho_{j}}|\varphi_{j}\rangle\langle\varphi_{j}|.
\end{equation}
In our derivation we will use several times the formula
\begin{equation}
\prod_{k=j}^{+\infty} {}_{k}\langle \beta_{k}|\alpha_{k}\rangle_{k}=
\prod_{k=j+1}^{+\infty} {}_{k}\langle \beta_{k}|\alpha_{k}\rangle_{k}\left(1-\frac{1}{2}\left(|\alpha_{j}|^2+|\beta_{j}|^2-2\alpha_{j}\beta_{j}^{\ast}\right)\tau+O(\tau^2)\right)
\end{equation}
following from 
\begin{equation}
_{k}\langle \beta_{k}|\alpha_{k}\rangle_{k}=1-\frac{1}{2}\left(|\alpha_{k}|^2+|\beta_{k}|^2-2\alpha_{k}\beta_{k}^{\ast}\right)\tau+O(\tau^2).
\end{equation}

Let us notice first that the conditional operator $\rho_{j+1}$ is given by the recurrence formula 
\begin{eqnarray}\label{filter2}
\rho_{j+1}&=&|c_{\alpha}|^2M_{\eta_{j+1}}^{\alpha_{j}}|\psi_{j}\rangle\langle\psi_{j}|M_{\eta_{j+1}}^{\alpha_{j} \dagger}+c_{\alpha}c_{\beta}^{\ast}\prod_{k=j+1}^{+\infty}\langle \beta_{k}|\alpha_{k}\rangle M_{\eta_{j+1}}^{\alpha_{j}}|\psi_{j}\rangle\langle\varphi_{j}|M_{\eta_{j+1}}^{\beta_{j}\dagger}
\nonumber\\
&&+c_{\alpha}^{\ast}c_{\beta}\prod_{k=j+1}^{+\infty}\langle \alpha_{k}|\beta_{k}\rangle M_{\eta_{j+1}}^{\beta_{j}}|\varphi_{j}\rangle\langle\psi_{j}|M_{\eta_{j+1}}^{\alpha_{j}\dagger}
+|c_{\beta}|^2M_{\eta_{j+1}}^{\beta_{j}}|\varphi_{j}\rangle\langle\varphi_{j}|M_{\eta_{j+1}}^{\beta_{j}\dagger},
\end{eqnarray}
where $\eta_{j+1}$ stands for the random variable having two possible values: $0$, $1$. Let us note that in order to determine $\tilde{\rho}_{j+1}$ we need to know the operators (\ref{oper1})--(\ref{oper4}) at the moment $j\tau$ and the result of the next measurement. When the result of the measurement is $0$, then we obtain from Eqs. (\ref{rec1}) and (\ref{rec2}) the following set of discrete equations 
\begin{eqnarray}
|\psi_{j+1}\rangle\langle\psi_{j+1}|&=&|\psi_{j}\rangle\langle\psi_{j}|
-i[H_{\mathcal{S}},|\psi_{j}\rangle\langle\psi_{j}|]\tau-\frac{1}{2}\left\{L^{\dagger}L,|\psi_{j}\rangle\langle\psi_{j}|\right\}\tau\nonumber\\
&&-|\psi_{j}\rangle\langle\psi_{j}|L\alpha_{j}^{\ast}\tau
-L^{\dagger}|\psi_{j}\rangle\langle\psi_{j}|\alpha_{j}\tau-|\psi_{j}\rangle\langle\psi_{j}||\alpha_{j}|^2\tau+O(\tau^2),
\end{eqnarray}
\begin{eqnarray}
|\psi_{j+1}\rangle\langle\varphi_{j+1}|&=&|\psi_{j}\rangle\langle\varphi_{j}|
\left(1-\frac{1}{2}\left(|\alpha_{j}|^2+|\beta_{j}|^2\right)\tau\right)\nonumber\\
&&-i[H_{\mathcal{S}},|\psi_{j}\rangle\langle\varphi_{j}|]\tau-\frac{1}{2}\left\{L^{\dagger}L,|\psi_{j}\rangle\langle\varphi_{j}|\right\}\tau\nonumber\\
&&-|\psi_{j}\rangle\langle\varphi_{j}|L\beta_{j}^{\ast}\tau
-L^{\dagger}|\psi_{j}\rangle\langle\varphi_{j}|\alpha_{j}\tau+O(\tau^2),
\end{eqnarray}
\begin{eqnarray}
|\varphi_{j+1}\rangle\langle\varphi_{j+1}|&=&|\varphi_{j}\rangle\langle\varphi_{j}|
-i[H_{\mathcal{S}},|\varphi_{j}\rangle\langle\varphi_{j}|]\tau-\frac{1}{2}\left\{L^{\dagger}L,|\varphi_{j}\rangle\langle\varphi_{j}|\right\}\tau\nonumber\\
&&-|\varphi_{j}\rangle\langle\varphi_{j}|L\beta_{j}^{\ast}\tau-L^{\dagger}|\varphi_{j}\rangle\langle\varphi_{j}|\beta_{j}\tau-|\varphi_{j}\rangle\langle\varphi_{j}||\beta_{j}|^2\tau+O(\tau^2).
\end{eqnarray}

The conditional probability of the outcome $0$ at the moment $(j+1)\tau$ when the {\it a posteriori} state of $\mathcal{S}$ at $j\tau$ was $\tilde{\rho}_{j}$ is defined as
\begin{equation}
p_{j+1}(0|\tilde{\rho}_{j})=\frac{\mathrm{Tr}{\rho}_{j+1}}{\mathrm{Tr}{\rho}_{j}},
\end{equation}
where $\rho_{j+1}$ is given by (\ref{filter2}) for $\eta_{j}=0$. Hence, we obtain the formula
\begin{equation}
p_{j+1}(0|\tilde{\rho}_{j})=1-\nu_{j}\tau+O(\tau^2),
\end{equation}
where
\begin{equation}\label{intensity}
\nu_{j}=|c_{\alpha}|^2\nu_{j}^{\alpha\alpha}+c_{\alpha}c_{\beta}^{\ast}\nu_{j}^{\alpha\beta}
+c_{\alpha}^{\ast}c_{\beta}\nu_{j}^{\beta\alpha}+|c_{\beta}|^2\nu_{j}^{\beta\beta},
\end{equation}
\begin{equation}
\nu_{j}^{\alpha\alpha}=\mathrm{Tr}\left[\left(L^{\dagger}L
+L\alpha_{j}^{\ast}+L^{\dagger}\alpha_{j}
+|\alpha_{j}|^2\right)\tilde{\rho}_{j}^{\alpha\alpha}\right],
\end{equation}
\begin{equation}
\nu_{j}^{\alpha\beta}=\mathrm{Tr}\left[\left(L^{\dagger}L
+L\beta_{j}^{\ast}+L^{\dagger}\alpha_{j}
+\alpha_{j}\beta_{j}^{\ast}\right)\tilde{\rho}_{j}^{\alpha\beta}\right],
\end{equation}
\begin{equation}
\nu_{j}^{\beta\alpha}=\mathrm{Tr}\left[\left(L^{\dagger}L
+L\alpha_{j}^{\ast}+L^{\dagger}\beta_{j}
+\alpha_{j}^{\ast}\beta_{j}\right)\tilde{\rho}_{j}^{\beta\alpha}\right],
\end{equation}
\begin{equation}
\nu_{j}^{\beta\beta}=\mathrm{Tr}\left[\left(L^{\dagger}L
+L\beta_{j}^{\ast}+L^{\dagger}\beta_{j}
+|\beta_{j}|^2\right)\tilde{\rho}_{j}^{\beta\beta}\right].
\end{equation}
Now, making use of the fact that
\begin{equation}
\frac{1}{\mathrm{Tr}\rho_{j+1}}=\frac{1}{\mathrm{Tr}\rho_{j}}
\left(1+\nu_{j} \tau+O(\tau^2)\right),
\end{equation}   
we obtain the set of difference equations
\begin{eqnarray}\label{rec0a}
\tilde{\rho}_{j+1}^{\alpha\alpha}-\tilde{\rho}_{j}^{\alpha\alpha}&=&\tilde{\rho}_{j}^{\alpha\alpha}\nu_{j}\tau
-i[H_{\mathcal{S}},\tilde{\rho}_{j}^{\alpha\alpha}]\tau-\frac{1}{2}\left\{L^{\dagger}L,\tilde{\rho}_{j}^{\alpha\alpha}\right\}\tau -\tilde{\rho}_{j}^{\alpha\alpha}L\alpha_{j}^{\ast}\tau \nonumber\\&&
-L^{\dagger}\tilde{\rho}_{j}^{\alpha\alpha}\alpha_{j}\tau-\tilde{\rho}_{j}^{\alpha\alpha}|\alpha_{j}|^2\tau+O(\tau^2),
\end{eqnarray}
\begin{eqnarray}\label{rec0b}
\tilde{\rho}_{j+1}^{\alpha\beta}-\tilde{\rho}_{j}^{\alpha\beta}&=&\tilde{\rho}_{j}^{\alpha\beta}\nu_{j}\tau
-i[H_{\mathcal{S}},\tilde{\rho}_{j}^{\alpha\beta}]\tau -\frac{1}{2}\left\{L^{\dagger}L,\tilde{\rho}_{j}^{\alpha\beta}\right\}\tau -\tilde{\rho}_{j}^{\alpha\beta}L\beta_{j}^{\ast}\tau \nonumber\\&&-L^{\dagger}\tilde{\rho}_{j}^{\alpha\beta}\alpha_{j}\tau-\tilde{\rho}_{j}^{\alpha\beta}\beta_{j}^{\ast}\alpha_{j}\tau+O(\tau^2),
\end{eqnarray}
\begin{eqnarray}\label{rec0c}
\tilde{\rho}_{j+1}^{\beta\beta}-\tilde{\rho}_{j}^{\beta\beta}&=&\tilde{\rho}_{j}^{\beta\beta}\nu_{j}\tau
-i[H_{\mathcal{S}},\tilde{\rho}_{j}^{\beta\beta}]\tau -\frac{1}{2}\left\{L^{\dagger}L,\tilde{\rho}_{j}^{\beta\beta}\right\}\tau -\tilde{\rho}_{j}^{\beta\beta}L\beta_{j}^{\ast}\tau \nonumber\\&&-L^{\dagger}\tilde{\rho}_{j}^{\beta\beta}\beta_{j}\tau -\tilde{\rho}_{j}^{\beta\beta}|\beta_{j}|^2\tau+O(\tau^2).
\end{eqnarray}
The equation for the operator $\tilde{\rho}_{j}^{\beta\alpha}$ one can get using the fact that $\tilde{\rho}_{j}^{\beta\alpha}=\left(\tilde{\rho}_{j}^{\alpha\beta}\right)^{\dagger}$. 

When the result of the measurement at the moment $(j+1)\tau$ is $1$, we get the following recurrence formulas
\begin{eqnarray}
\lefteqn{|\psi_{j+1}\rangle\langle\psi_{j+1}|=\left(L|\psi_{j}\rangle\langle\psi_{j}|L^{\dagger}+L|\psi_{j}\rangle\langle\psi_{j}|
	\alpha_{j}^{\ast}\right.}\nonumber\\
&&\left.+|\psi_{j}\rangle\langle\psi_{j}|L^{\dagger}\alpha_{j} +|\psi_{j}\rangle\langle\psi_{j}||\alpha_{j}|^2\right)\tau+O(\tau^2),
\end{eqnarray}
\begin{eqnarray}
\lefteqn{|\psi_{j+1}\rangle\langle\varphi_{j+1}|=\left(L|\psi_{j}\rangle\langle\varphi_{j}|L^{\dagger}+
	L|\psi_{j}\rangle\langle\varphi_{j}|\beta_{j}^{\ast}\right.}\nonumber\\
&&\left.+|\psi_{j}\rangle\langle\varphi_{j}|
L^{\dagger}\alpha_{j} +|\psi_{j}\rangle\langle\varphi_{j}|\alpha_{j}\beta_{j}^{\ast}\right)\tau+O(\tau^2),
\end{eqnarray}
\begin{eqnarray}
\lefteqn{|\varphi_{j+1}\rangle\langle\varphi_{j+1}|=\left(L|\varphi_{j}\rangle\langle\varphi_{j}|L^{\dagger}+
	L|\varphi_{j}\rangle\langle\varphi_{j}|\beta_{j}^{\ast}\right.}\nonumber\\
&&\left.+|\varphi_{j}\rangle\langle\varphi_{j}|
L^{\dagger}\beta_{j} +|\varphi_{j}\rangle\langle\varphi_{j}||\beta_{j}|^2\right)\tau+O(\tau^2).
\end{eqnarray}

The conditional probability of the outcome $1$ at the moment $(j+1)\tau$ when the {\it a posteriori} state of $\mathcal{S}$ at the moment $j\tau$ was $\tilde{\rho}_{j}$ is defined by
\begin{equation}
p_{j+1}(1|\tilde{\rho}_{j})=\frac{\mathrm{Tr}{\rho}_{j+1}}{\mathrm{Tr}{\rho}_{j}},
\end{equation}
where $\rho_{j+1}$ is given by (\ref{filter2}) with $\eta_{j}=1$. One can check that  
\begin{equation}
p_{j+1}(1|\tilde{\rho}_{j})=\nu_{j}\tau+O(\tau^2),
\end{equation} 
where the conditional intensity $\nu_{j}$ is defined by (\ref{intensity}). So for the result $1$ we find that
\begin{eqnarray}\label{rec1a}
\tilde{\rho}_{j+1}^{\alpha\alpha}=\frac{1}{\nu_{j}}\left(L\tilde{\rho}_{j}^{\alpha\alpha}L^{\dagger}+L\tilde{\rho}_{j}^{\alpha\alpha}
\alpha_{j}^{\ast}+\tilde{\rho}_{j}^{\alpha\alpha}L^{\dagger}\alpha_{j} +\tilde{\rho}_{j}^{\alpha\alpha}|\alpha_{j}|^2\right)+O(\tau),
\end{eqnarray}
\begin{eqnarray}\label{rec1b}
\tilde{\rho}_{j+1}^{\alpha\beta}=\frac{1}{\nu_{j}}\left(L\tilde{\rho}_{j}^{\alpha\beta}L^{\dagger}+
L\tilde{\rho}_{j}^{\alpha\beta}\beta_{j}^{\ast}+\tilde{\rho}_{j}^{\alpha\beta}
L^{\dagger}\alpha_{j} +\tilde{\rho}_{j}^{\alpha\beta}\alpha_{j}\beta_{j}^{\ast}\right)+O(\tau),
\end{eqnarray}
\begin{eqnarray}\label{rec1c}
\tilde{\rho}_{j+1}^{\beta\beta}=\frac{1}{\nu_{j}}\left(L\tilde{\rho}_{j}^{\beta\beta}L^{\dagger}+
L\tilde{\rho}_{j}^{\beta\beta}\beta_{j}^{\ast}+\tilde{\rho}_{j}^{\beta\beta}
L^{\dagger}\beta_{j} +\tilde{\rho}_{j}^{\beta\beta}|\beta_{j}|^2\right)+O(\tau).
\end{eqnarray}

Let us introduce now the stochastic discrete process
\begin{equation}
n_{j}=\sum_{k=0}^{j}\eta_{k},
\end{equation}
with the increment
\begin{equation}
\Delta n_{j}=n_{j+1}-n_{j}.
\end{equation}
One can check that the conditional expectation
\begin{equation}
\mathbbm{E}[\Delta n_{j}|\tilde{\rho}_{j}]=\nu_{j}\tau+O(\tau^2).
\end{equation} 

Finally, by combining Eqs. (\ref{rec0a})--(\ref{rec0c}) with Eqs. (\ref{rec1a})--(\ref{rec1c}), we obtain the set of stochastic difference equations
\begin{eqnarray}\label{filter3}
\tilde{\rho}_{j+1}^{\alpha\alpha}-\tilde{\rho}_{j}^{\alpha\alpha}&=&\mathcal{L}\rho_{j}^{\alpha\alpha}\tau+[\tilde{\rho}_{j}^{\alpha\alpha},L^{\dagger}]\alpha_{j}\tau
+[L,\tilde{\rho}_{j}^{\alpha\alpha}]\alpha^{\ast}_{j}\tau\nonumber\\
&&+\bigg\{\frac{1}{\nu_{j}}\left(
L\tilde{\rho}_{j}^{\alpha\alpha}L^{\dagger}+\tilde{\rho}_{j}^{\alpha\alpha}L^{\dagger}\alpha_{j}
+L\tilde{\rho}_{j}^{\alpha\alpha}\alpha^{\ast}_{j}
\right.\nonumber\\
&&\left.+\tilde{\rho}_{j}^{\alpha\alpha}|\alpha_{j}|^2\right)
-\tilde{\rho}_{j}^{\alpha\alpha}\bigg\}\left(\Delta n_{j}-\nu_{j}\tau\right),
\end{eqnarray}
\begin{eqnarray}\label{filter4}
\tilde{\rho}_{j+1}^{\alpha\beta}-\tilde{\rho}_{j}^{\alpha\beta}&=&\mathcal{L}\rho_{j}^{\beta\beta}\tau+[\tilde{\rho}_{j}^{\alpha\beta},L^{\dagger}]\alpha_{j}\tau
+[L,\tilde{\rho}_{j}^{\alpha\beta}]\beta^{\ast}_{j}\tau
\nonumber\\
&&+\bigg\{\frac{1}{\nu_{j}}\left(
L\tilde{\rho}_{j}^{\alpha\beta}L^{\dagger}+\tilde{\rho}_{j}^{\alpha\beta}L^{\dagger}\alpha_{j}
+L\tilde{\rho}_{j}^{\alpha\beta}\beta^{\ast}_{j}\right.\nonumber\\
&&\left.
+\tilde{\rho}_{j}^{\alpha\beta}\beta_{j}^{\ast}\alpha_{j}\right)
-\tilde{\rho}_{j}^{\alpha\beta}\bigg\}\left(\Delta n_{j}-\nu_{j}\tau\right)
\end{eqnarray} 
\begin{eqnarray}\label{filter5}
\tilde{\rho}_{j+1}^{\beta\beta}-\tilde{\rho}_{j}^{\beta\beta}&=& 
\mathcal{L}\rho_{j}^{\beta\alpha}\tau+[\tilde{\rho}_{j}^{\beta\beta},L^{\dagger}]\beta_{j}\tau
+[L,\tilde{\rho}_{j}^{\beta\beta}]\beta^{\ast}_{j}\tau\nonumber\\
&&+\bigg\{\frac{1}{\nu_{j}}\left(
L\tilde{\rho}_{j}^{\beta\beta}L^{\dagger}+\tilde{\rho}_{j}^{\beta\beta} L^{\dagger}\beta_{j}+L\tilde{\rho}_{j}^{\beta\beta}\beta^{\ast}_{j}\right.\nonumber\\
&&\left.
+\tilde{\rho}_{j}^{\beta\beta}|\beta_{t}|^2\right)
-\tilde{\rho}_{j}^{\beta\beta}\bigg\}\left(\Delta n_{j}-\nu_{j}\tau\right),
\end{eqnarray}
where
\begin{equation}\label{superop}
\mathcal{L}\rho=-i[H_{\mathcal{S}},\rho]
-\frac{1}{2}\left\{L^{\dagger}L,\rho\right\}
+L\rho L^{\dagger}
\end{equation}
and the initial condition
$\tilde{\rho}_{0}^{\alpha\alpha}=\tilde{\rho}_{0}^{\beta\beta}=|\psi\rangle\langle\psi|$, $\tilde{\rho}_{0}^{\alpha\beta}=\langle\beta|\alpha\rangle|\psi\rangle\langle\psi|$. We dropped here all terms that do not contribute to the continuous time limit when $\tau\to dt$. Note that when $\Delta n_{j}$ is equal to $0$, then Eqs. (\ref{filter3})--(\ref{filter5}) reduce to Eqs. (\ref{rec0a})--(\ref{rec0c}), and when $\Delta n_{j}$ is equal to $1$, then all the terms proportional to $\tau$ in Eqs. (\ref{filter3})--(\ref{filter5}) are negligible and we obtain the formulas (\ref{rec1a})--(\ref{rec1c}).

Let us notice that to get the continuous in time evolution of $\mathcal{S}$ we fix time $t=j\tau$ such that when $j\to+\infty$ we have $\tau\to 0$. Of course, we take $t$ fixed but arbitrary. Thus in the continuous time limit we get from (\ref{filter3})--(\ref{filter5}) the set of the stochastic differential equations of the form
\begin{eqnarray}\label{filter6}
d\tilde{\rho}_{t}^{\alpha\alpha}&=&\mathcal{L}\rho_{t}^{\alpha\alpha}dt+[\tilde{\rho}_{t}^{\alpha\alpha},L^{\dagger}]\alpha_{t}dt
+[L,\tilde{\rho}_{t}^{\alpha\alpha}]\alpha^{\ast}_{t}dt
\nonumber\\
&&+\bigg\{\frac{1}{\nu_{j}}\left(
L\tilde{\rho}_{t}^{\alpha\alpha}L^{\dagger}+\tilde{\rho}_{t}^{\alpha\alpha}L^{\dagger}\alpha_{t}
+L\tilde{\rho}_{t}^{\alpha\alpha}\alpha^{\ast}_{t}
\right.\nonumber\\
&&\left.+\tilde{\rho}_{t}^{\alpha\alpha}|\alpha_{t}|^2\right)
-\tilde{\rho}_{t}^{\alpha\alpha}\bigg\}\left(dn_{t}-\nu_{t}dt\right),
\end{eqnarray}
\begin{eqnarray}\label{filter7}
d\tilde{\rho}_{t}^{\alpha\beta}&=&\mathcal{L}\rho_{t}^{\alpha\beta}dt
+[\tilde{\rho}_{t}^{\alpha\beta},L^{\dagger}]\alpha_{t}dt
+[L,\tilde{\rho}_{t}^{\alpha\beta}]\beta^{\ast}_{t}dt
\nonumber\\
&&+\bigg\{\frac{1}{\nu_{t}}\left(
L\tilde{\rho}_{t}^{\alpha\beta}L^{\dagger}
+\tilde{\rho}_{t}^{\alpha\beta}L^{\dagger}\alpha_{t}
+L\tilde{\rho}_{t}^{\alpha\beta}\beta^{\ast}_{t}
\right.\nonumber\\
&&\left.+\tilde{\rho}_{t}^{\alpha\beta}\beta_{t}^{\ast}\alpha_{t}\right)
-\tilde{\rho}_{t}^{\alpha\beta}\bigg\}\left(dn_{t}-\nu_{t}dt\right)
\end{eqnarray} 
\begin{eqnarray}\label{filter8}
d\tilde{\rho}_{t}^{\beta\beta}&=& \mathcal{L}\rho_{t}^{\beta\beta}dt+[\tilde{\rho}_{t}^{\beta\beta},L^{\dagger}]\beta_{t}dt
+[L,\tilde{\rho}_{t}^{\beta\beta}]\beta^{\ast}_{t}dt
\nonumber\\
&&+\bigg\{\frac{1}{\nu_{t}}\left(
L\tilde{\rho}_{t}^{\beta\beta}L^{\dagger}+\tilde{\rho}_{t}^{\beta\beta} L^{\dagger}\beta_{t}+L\tilde{\rho}_{t}^{\beta\beta}\beta^{\ast}_{j}
\right.\nonumber\\
&&\left.+\tilde{\rho}_{t}^{\beta\beta}|\beta_{t}|^2\right)
-\tilde{\rho}_{t}^{\beta\beta}\bigg\}\left(dn_{t}-\nu_{t}dt\right)
\end{eqnarray} 
and initially $\tilde{\rho}_{0}^{\alpha\alpha}=\tilde{\rho}_{0}^{\beta\beta}=|\psi\rangle\langle\psi|$, $\tilde{\rho}_{0}^{\alpha\beta}=\langle \beta|\alpha\rangle|\psi\rangle\langle\psi|$. 
The stochastic process $n_{t}$ is defined as the continuous limit of the discrete process $n_{j}$. The It\^{o} table for $dn_{t}$ is $\left(dn_{t}\right)^2=dn_{t}$ (we can measure at most one photon in the interval of length $dt$) and  $\mathbbm{E}\left[dn_{t}|\tilde{\rho}_{t}\right]=\nu_{t}dt$, where
\begin{equation}
\nu_{t}=|c_{\alpha}|^2\nu_{t}^{\alpha\alpha}+c_{\alpha}c_{\beta}^{\ast}\nu_{t}^{\alpha\beta}
+c_{\alpha}^{\ast}c_{\beta}\nu_{t}^{\beta\alpha}+|c_{\beta}|^2\nu_{t}^{\beta\beta},
\end{equation}
\begin{equation}
\nu_{t}^{\alpha\alpha}=\mathrm{Tr}\left[\left(L^{\dagger}L
+L\alpha_{t}^{\ast}+L^{\dagger}\alpha_{t}
+|\alpha_{t}|^2\right)\tilde{\rho}_{t}^{\alpha\alpha}\right],
\end{equation}
\begin{equation}
\nu_{t}^{\alpha\beta}=\mathrm{Tr}\left[\left(L^{\dagger}L
+L\beta_{t}^{\ast}+L^{\dagger}\alpha_{t}
+\alpha_{t}\beta_{t}^{\ast}\right)\tilde{\rho}_{t}^{\alpha\beta}\right],
\end{equation}
\begin{equation}
\nu_{t}^{\beta\alpha}=\mathrm{Tr}\left[\left(L^{\dagger}L
+L\alpha_{t}^{\ast}+L^{\dagger}\beta_{t}
+\alpha_{t}^{\ast}\beta_{t}\right)\tilde{\rho}_{t}^{\beta\alpha}\right],
\end{equation}
\begin{equation}
\nu_{t}^{\beta\beta}=\mathrm{Tr}\left[\left(L^{\dagger}L
+L\beta_{t}^{\ast}+L^{\dagger}\beta_{t}
+|\beta_{t}|^2\right)\tilde{\rho}_{t}^{\beta\beta}\right].
\end{equation}
Moreover, the complex functions $\alpha_{t}$ and $\beta_{t}$ satisfy the conditions
\begin{eqnarray}
\int_{0}^{+\infty}|\alpha_{t}|^2dt<+\infty,\;\;\;\;\int_{0}^{+\infty}|\beta_{t}|^2dt<+\infty,
\end{eqnarray}
and
\begin{equation}
\langle \beta|\alpha\rangle=\exp\left\{-\frac{1}{2}\int_{0}^{+\infty}\left(|\alpha_{t}|^2+|\beta_{t}|^2-2\alpha_{t}\beta^{\ast}_{t}\right)dt\right\}.
\end{equation} 
Thus, the {\it a posteriori} state of $\mathcal{S}$ is given as
\begin{equation}
\tilde{\rho}_{t}=
|c_{\alpha}|^2\tilde{\rho}_{t}^{\alpha\alpha}+c_{\alpha}c_{\beta}^{\ast}\tilde{\rho}_{t}^{\alpha\beta}
+c_{\alpha}^{\ast}c_{\beta}\tilde{\rho}_{t}^{\beta\alpha}
+|c_{\beta}|^2\tilde{\rho}_{t}^{\beta\beta},
\end{equation}
where the conditional operators $\tilde{\rho}_{t}^{\alpha\alpha}$, $\tilde{\rho}_{t}^{\alpha\beta}$, $\tilde{\rho}_{t}^{\beta\beta}$ satisfy Eqs. (\ref{filter6})-(\ref{filter8}), and $\tilde{\rho}_{t}^{\beta\alpha}=\left(\tilde{\rho}_{t}^{\alpha\beta}\right)^{\dagger}$. One can check that $\mathrm{Tr}\tilde{\rho}_{t}=1$ for any $t\geq 0$. The equations (\ref{filter6})-(\ref{filter8}) agree with the stochastic master equations derived in \cite{GJN12b} (see Sec. IV in \cite{GJN12b}).

When we take an average of $\tilde{\rho}_{t}$ over all realizations of the stochastic process $n_{t}$ (all possible outcomes) then we get the {\it a priori} evolution of the system $\mathcal{S}$. One can check that the {\it a priori} state of $\mathcal{S}$ is described by  
\begin{equation}
{\varrho}_{t}=
|c_{\alpha}|^2{\varrho}_{t}^{\alpha\alpha}+c_{\alpha}c_{\beta}^{\ast}{\varrho}_{j}^{\alpha\beta}
+c_{\alpha}^{\ast}c_{\beta}{\varrho}_{t}^{\beta\alpha}
+|c_{\beta}|^2{\varrho}_{t}^{\beta\beta},
\end{equation}
where the operators ${\varrho}_{t}^{\alpha\alpha}$, ${\varrho}_{t}^{\alpha\beta}$,
${\varrho}_{t}^{\beta\beta}$ satisfy the differential equations
\begin{eqnarray}\label{master1}
\dot{{\varrho}}_{t}^{\alpha\alpha}&=&\mathcal{L}\varrho_{t}^{\alpha\alpha}+[{\varrho}_{t}^{\alpha\alpha},L^{\dagger}]\alpha_{t}
+[L,{\varrho}_{t}^{\alpha\alpha}]\alpha^{\ast}_{t},
\end{eqnarray}
\begin{eqnarray}
\dot{{\varrho}}_{t}^{\alpha\beta}&=&\mathcal{L}\varrho_{t}^{\alpha\beta}+[{\varrho}_{t}^{\alpha\beta},L^{\dagger}]\alpha_{t}
+[L,{\varrho}_{t}^{\alpha\beta}]\beta^{\ast}_{t},
\end{eqnarray} 
\begin{eqnarray}\label{master3}
\dot{{\varrho}}_{t}^{\beta\beta}&=& \mathcal{L}\rho_{t}^{\beta\beta}+[{\rho}_{t}^{\beta\beta},L^{\dagger}]\beta_{t}
+[L,{\rho}_{t}^{\beta\beta}]\beta^{\ast}_{t},
\end{eqnarray} 
where $\mathcal{L}$ acts as (\ref{superop}). The initial condition is 
${\varrho}_{0}^{\alpha\alpha}={\varrho}_{0}^{\beta\beta}=|\psi\rangle\langle\psi|$, ${\varrho}_{0}^{\alpha\beta}=\langle \beta|\alpha\rangle|\psi\rangle\langle\psi|$, and ${\varrho}_{t}^{\beta\alpha}=\left({\varrho}_{t}^{\alpha\beta}\right)^{\dagger}$. One can easily check that $\mathrm{Tr}\varrho_{t}^{\alpha\alpha}=\mathrm{Tr}\varrho_{t}^{\beta\beta}=1$, $\mathrm{Tr}\varrho_{t}^{\alpha\beta}=\langle \beta|\alpha\rangle$, and $\mathrm{Tr}\varrho_{t}^{\beta\alpha}=\langle \alpha|\beta\rangle$ for any $t\geq 0$.

\subsection{Homodyne detection}

\begin{Theorem}\label{TH-4} The conditional state of $\mathcal{S}$ and the part of the environment which has not interacted with $\mathcal{S}$ up to $j\tau$ for the initial state (\ref{ini2}) and the measurement of (\ref{obs2}) at the moment $j\tau$ is given by
	\begin{equation}\label{cond5}
	|\tilde{\Psi}_{j}\rangle = \frac{|\Psi_{j}\rangle}{\sqrt{\langle\Psi_{j}|\Psi_{j}\rangle}},
	\end{equation}
	where
	\begin{equation}\label{cond6}
	|\Psi_{j}\rangle=c_{\alpha}\bigotimes_{k=j}^{+\infty}|\alpha_{k}\rangle_{k}\otimes|\psi_{j}\rangle+
	c_{\beta}\bigotimes_{k=j}^{+\infty}|\beta_{k}\rangle_{k}\otimes|\varphi_{j}\rangle.
	\end{equation}
	The conditional vectors $|\psi_{j}\rangle$, $|\varphi_{j}\rangle$ from $\mathcal{H}_{\mathcal{S}}$ in (\ref{cond4}) are given by the recurrence formulas
	\begin{equation}\label{rec3}
	|\psi_{j+1}\rangle=R_{\zeta_{j+1}}^{\alpha_{j}}|\psi_{j}\rangle,
	\end{equation}
	\begin{equation}\label{rec4}
	|\varphi_{j+1}\rangle=R_{\zeta_{j+1}}^{\beta_{j}}|\varphi_{j}\rangle,
	\end{equation}
	where $\zeta_{j+1}$ stands for a random variable describing the $(j+1)$-th output of (\ref{obs2}), and
	\begin{eqnarray}
	R_{\zeta_{j+1}}^{\alpha_{j}}&=&\frac{1}{\sqrt{2}}\bigg[\mathbbm{1}_{S}-\left(iH_{\mathcal{S}}+\frac{1}{2}L^{\dagger}L+L^{\dagger}\alpha_{j}+\frac{|\alpha_{j}|^2}{2}\right)\tau\nonumber\\&&+(L+\alpha_{j}) \zeta_{j+1}\sqrt{\tau}+O\left(\tau^{3/2}\right)\bigg],
	\end{eqnarray}
	\begin{eqnarray}
	R_{\zeta_{j+1}}^{\beta_{j}}&=&\frac{1}{\sqrt{2}}\bigg[\mathbbm{1}_{S}-\left(iH_{\mathcal{S}}+\frac{1}{2}L^{\dagger}L+L^{\dagger}\beta_{j}+\frac{|\beta_{j}|^2}{2}\right)\tau\nonumber\\&&+(L+\beta_{j}) \zeta_{j+1}\sqrt{\tau}+O\left(\tau^{3/2}\right)\bigg],
	\end{eqnarray}
and initially we have $|\psi_{0}\rangle=|\varphi_{0}\rangle=|\psi\rangle$. 
\end{Theorem}

{\it Proof.} To prove Theorem (\ref{TH-4}) we use the result of Sec. III.B and the linearity of the evolution equation for the total system.

Clearly, the conditional state of $\mathcal{S}$ at the moment $j\tau$ has the form (\ref{Filter}). We start derivation of the filtering equations for the stochastic operators (\ref{oper1})--(\ref{oper4}) from writing down the recursive formulas   
\begin{eqnarray}
2|\psi_{j+1}\rangle\langle\psi_{j+1}|&=&|\psi_{j}\rangle\langle\psi_{j}|
+\mathcal{L}|\psi_{j}\rangle\langle\psi_{j}|\tau\nonumber\\&&
+[|\psi_{j}\rangle\langle\psi_{j}|,L^{\dagger}]\alpha_{j}\tau
+[L,|\psi_{j}\rangle\langle\psi_{j}|]\alpha_{j}^{\ast}\tau\nonumber\\
&&\!+\!\left[\left(L+\alpha_{j}\right)|\psi_{j}\rangle\langle\psi_{j}|
+|\psi_{j}\rangle\langle\psi_{j}|\left(L^{\dagger}+\alpha_{j}^{\ast}\right)\right]\zeta_{j+1}\sqrt{\tau},
\end{eqnarray}
\begin{eqnarray}
2|\psi_{j+1}\rangle\langle\varphi_{j+1}|&=&|\psi_{j}\rangle\langle\varphi_{j}|
\left(1-\frac{1}{2}\left(|\alpha_{j}|^2+|\beta_{j}|^2-2\beta_{j}^{\ast}\alpha_{j}\right)\tau\right)\nonumber\\
&&+\mathcal{L}|\psi_{j+1}\rangle\langle\varphi_{j+1}|\tau\nonumber\\
&&+[|\psi_{j}\rangle\langle\varphi_{j}|,L^{\dagger}]\alpha_{j}\tau
+[L,|\psi_{j}\rangle\langle\varphi_{j}|]\beta_{j}^{\ast}\tau
\nonumber\\
&&+\left[\left(L+\alpha_{j}\right)|\psi_{j}\rangle\langle\varphi_{j}|+|\psi_{j}\rangle\langle\varphi_{j}|\left(L^{\dagger}+\beta_{j}^{\ast}\right)\right] \zeta_{j+1}\sqrt{\tau},
\end{eqnarray}
\begin{eqnarray}
2|\varphi_{j+1}\rangle\langle\varphi_{j+1}|&=&|\varphi_{j}\rangle\langle\varphi_{j}|+\mathcal{L}|\varphi_{j}\rangle\langle\varphi_{j}|\tau \nonumber\\&&+[|\varphi_{j}\rangle\langle\varphi_{j}|,L]\beta_{j}dt+[L,|\varphi_{j}\rangle\langle\varphi_{j}|]\beta_{j}^{\ast}\tau
\nonumber\\
&&+\left[\left(L+\beta_{j}\right)|\varphi_{j}\rangle\langle\varphi_{j}|
+|\varphi_{j}\rangle\langle\varphi_{j}|\left(L^{\dagger}+\beta_{j}^{\ast}\right)\right] \zeta_{j+1}\sqrt{\tau},
\end{eqnarray}

We can readily deduce that the conditional probability of the result $\zeta_{j+1}$ at the moment $(j+1)\tau$ when the conditional state of $\mathcal{S}$ is $\tilde{\rho}_{j}$ at the time $j\tau$ is given by
\begin{equation}
p_{j+1}(\zeta_{j+1}|\tilde{\rho}_{j})=
\frac{1}{2}\left(1+\mu_{j}\zeta_{j+1}\sqrt{\tau}\right)+O(\tau^{3/2}),
\end{equation}
where
\begin{equation}
\mu_{j}=|c_{\alpha}|^2\mu_{j}^{\alpha\alpha}+c_{\alpha}c_{\beta}^{\ast}\mu_{j}^{\alpha\beta}
+c_{\alpha}^{\ast}c_{\beta}\mu_{j}^{\beta\alpha}+|c_{\beta}|^2\mu_{j}^{\beta\beta}
\end{equation}
and
\begin{equation}\label{intensity21}
\mu_{j}^{\alpha\alpha}=
\mathrm{Tr}\left[\left(L+L^{\dagger}+\alpha_{j}+\alpha_{j}^{\ast}\right)\tilde{\rho}_{j}^{\alpha\alpha}\right],
\end{equation}
\begin{equation}\label{intensity22}
\mu_{j}^{\alpha\beta}=\mathrm{Tr}\left[\left(L+L^{\dagger}
+\alpha_{j}+\beta_{j}^{\ast}\right)\tilde{\rho}_{j}^{\alpha\beta}\right],
\end{equation}
\begin{equation}\label{intensity23}
\mu_{j}^{\beta\alpha}=\mathrm{Tr}\left[\left(L+L^{\dagger}
+\beta_{j}+\alpha_{j}^{\ast}\right)\tilde{\rho}_{j}^{\beta\alpha}\right],
\end{equation}
\begin{equation}\label{intensity24}
\mu_{j}^{\beta\beta}=\mathrm{Tr}\left[\left(L+L^{\dagger}
+\beta_{j}+\beta_{j}^{\ast}\right)\tilde{\rho}_{j}^{\beta\beta}\right].
\end{equation}
Thus for the discrete stochastic process $\zeta_{j}$ we obtain the conditional mean values 
\begin{equation}
\mathbbm{E}[\zeta_{j+1}|\tilde{\rho}_{j}]=\mu_{j}\sqrt{\tau}+O(\tau^{3/2}),
\end{equation}
\begin{equation}
\mathbbm{E}[\zeta_{j+1}^2|\tilde{\rho}_{j}]=1+O(\tau^{3/2}).
\end{equation}
Let us introduce now the stochastic process
\begin{equation}
q_{j}=\sqrt{\tau}\sum_{k=1}^{j}\zeta_{k}
\end{equation}
One can easily check that $\mathbbm{E}[\Delta q_{j}=q_{j+1}-q_{j}|\tilde{\rho}_{j}]=\mu_{j}\tau+O(\tau^{3/2})$. 

Now, taking into account that
\begin{equation}
\frac{1}{\mathrm{Tr}\rho_{j+1}}=\frac{2}{\mathrm{Tr}\rho_{j}}
\left(1-\mu_{j}\zeta_{j+1}\sqrt{\tau}+\mu_{j}^2\tau\right)
\end{equation}
after some algebra we find the set of the stochastic difference equations 
\begin{eqnarray}
\tilde{\rho}_{j+1}^{\alpha\alpha}-\tilde{\rho}_{j}^{\alpha\alpha}&=&\mathcal{L}\tilde{\rho}_{j}^{\alpha\alpha}\tau+[\tilde{\rho}_{j}^{\alpha\alpha},L^{\dagger}]\alpha_{j}\tau
+[L,\tilde{\rho}_{j}^{\alpha\alpha}]\alpha^{\ast}_{j}\tau
\nonumber\\
&&+\left[\left(L+\alpha_{j}\right)\tilde{\rho}_{j}^{\alpha\alpha}+
\tilde{\rho}_{j}^{\alpha\alpha}(L^{\dagger}+\alpha_{j}^{\ast})
-\mu_{j}\tilde{\rho}_{j}^{\alpha\alpha}\right]
\left(\Delta q_{j}-\mu_{j}\tau\right),
\end{eqnarray} 
\begin{eqnarray}
\tilde{\rho}_{j+1}^{\alpha\beta}-\tilde{\rho}_{j}^{\alpha\beta}&=&\mathcal{L}\tilde{\rho}_{j}^{\alpha\beta}\tau+[\tilde{\rho}_{j}^{\alpha\beta},L^{\dagger}]\alpha_{j}\tau
+[L,\tilde{\rho}_{j}^{\alpha\beta}]\beta^{\ast}_{j}\tau
\nonumber\\
&&+\left[\left(L+\alpha_{j}\right)\tilde{\rho}_{j}^{\alpha\beta}
+\tilde{\rho}_{j}^{\alpha\beta}(L^{\dagger}+\beta_{j}^{\ast})
-\mu_{j}\tilde{\rho}_{j}^{\alpha\beta}\right]\left(\Delta q_{j}-\mu_{j}\tau\right),
\end{eqnarray}
\begin{eqnarray}
\tilde{\rho}_{j+1}^{\beta\beta}-\tilde{\rho}_{j}^{\beta\beta}&=& 
\mathcal{L}\tilde{\rho}_{j}^{\beta\beta}\tau+[\tilde{\rho}_{j}^{\beta\beta},L^{\dagger}]\beta_{j}\tau
+[L,\tilde{\rho}_{j}^{\beta\beta}]\beta^{\ast}_{j}\tau
\nonumber\\
&&+\left[\left(L+\beta_{j}\right)\tilde{\rho}_{j}^{\beta\beta}
+\tilde{\rho}_{j}^{\beta\beta}\left(L^{\dagger}+\beta_{j}^{\ast}\right)
-\mu_{j}\tilde{\rho}_{j}^{\beta\beta}\right]\left(\Delta q_{j}-\mu_{j}\tau\right)
\end{eqnarray}
with the initial conditions $\tilde{\rho}_{0}^{\alpha\alpha}=|\psi\rangle\langle\psi|$, $\tilde{\rho}_{0}^{\beta\beta}=|\psi\rangle\langle\psi|$, $\tilde{\rho}_{0}^{\alpha\beta}=\langle\beta|\alpha\rangle|\psi\rangle\langle\psi|$. 
Then in the continuous time limit we have
\begin{eqnarray}
d\tilde{\rho}_{t}^{\alpha\alpha}&=&\mathcal{L}\rho_{t}^{\alpha\alpha}dt+[\tilde{\rho}_{t}^{\alpha\alpha},L^{\dagger}]\alpha_{t}dt
+[L,\tilde{\rho}_{t}^{\alpha\alpha}]\alpha^{\ast}_{t}dt
\nonumber\\
&&+\left[\left(L+\alpha_{t}\right)\tilde{\rho}_{t}^{\alpha\alpha}
+\tilde{\rho}_{t}^{\alpha\alpha}\left(L^{\dagger}+\alpha_{t}^{\ast}\right)
-\mu_{t}\tilde{\rho}_{t}^{\alpha\alpha}\right]\left(dq_{t}-\mu_{t}dt\right)
\end{eqnarray} 
\begin{eqnarray}
d\tilde{\rho}_{t}^{\alpha\beta}&=&\mathcal{L}\rho_{t}^{\alpha\beta}dt
+[\tilde{\rho}_{t}^{\alpha\beta},L^{\dagger}]\alpha_{t}dt
+[L,\tilde{\rho}_{t}^{\alpha\beta}]\beta^{\ast}_{t}dt
\nonumber\\
&&+\left[\left(L+\alpha_{t}\right)\tilde{\rho}_{t}^{\alpha\beta}
+\tilde{\rho}_{t}^{\alpha\beta}\left(L^{\dagger}+\beta_{t}^{\ast}\right)
-\mu_{t}\tilde{\rho}_{t}^{\alpha\beta}\right]\left(dq_{t}-\mu_{t}dt\right),
\end{eqnarray}
\begin{eqnarray}
d\tilde{\rho}_{t}^{\beta\beta}&=& \mathcal{L}\rho_{t}^{\beta\beta}dt
+[\tilde{\rho}_{t}^{\beta\beta},L^{\dagger}]\beta_{t}dt
+[L,\tilde{\rho}_{t}^{\beta\beta}]\beta^{\ast}_{t}dt
\nonumber\\
&&+\left[\left(L+\beta_{t}\right)\tilde{\rho}_{t}^{\beta\beta}
+\tilde{\rho}_{t}^{\beta\beta}\left(L^{\dagger}+\beta_{t}^{\ast}\right)
-\mu_{t}\tilde{\rho}_{t}^{\beta\beta}\right]\left(dq_{t}-\mu_{t}dt\right)
\end{eqnarray}
where
\begin{equation}
\mu_{t}=|c_{\alpha}|^2\mu_{t}^{\alpha\alpha}+c_{\alpha}c_{\beta}^{\ast}\mu_{t}^{\alpha\beta}
+c_{\alpha}^{\ast}c_{\beta}\mu_{t}^{\beta\alpha}+|c_{\beta}|^2\mu_{t}^{\beta\beta}
\end{equation}
and
\begin{equation}
\mu_{t}^{\alpha\alpha}=
\mathrm{Tr}\left[\left(L+L^{\dagger}+\alpha_{t}+\alpha_{t}^{\ast}\right)\tilde{\rho}_{t}^{\alpha\alpha}\right],
\end{equation}
\begin{equation}
\mu_{t}^{\alpha\beta}=\mathrm{Tr}\left[\left(L+L^{\dagger}
+\alpha_{t}+\beta_{t}^{\ast}\right)\tilde{\rho}_{t}^{\alpha\beta}\right],
\end{equation}
\begin{equation}
\mu_{t}^{\beta\alpha}=\mathrm{Tr}\left[\left(L+L^{\dagger}
+\beta_{t}+\alpha_{t}^{\ast}\right)\tilde{\rho}_{t}^{\beta\alpha}\right],
\end{equation}
\begin{equation}
\mu_{t}^{\beta\beta}=\mathrm{Tr}\left[\left(L+L^{\dagger}
+\beta_{t}+\beta_{t}^{\ast}\right)\tilde{\rho}_{t}^{\beta\beta}\right].
\end{equation}
The process $q_{j}$ in the limit $\tau\to 0$ converges to the stochastic process $q_{t}$ with the conditional probability $\mathbbm{E}[d q_{t}=q_{t+dt}-q_{t}|\tilde{\rho}_{t}]=\mu_{t}dt$. 

\section{An example: a cavity mode}
About the emergence of collision model in quantum optics one can read, for instance, in \cite{C17, FTVRS18}. To derive the discrete model of repeated interactions and measurements one starts from description of interaction of a quantum system with a Bose field propagating in only one direction, making the rotating wave approximation and taking the flat spectrum of the field. Then one passes to the interaction picture with respect to the free dynamics of the field and takes the Hamiltonian of the field in the frequency domain with the lower limit of integration extended to $-\infty$. The time coarse-graining model arises from division of the field into some probe segments. From the form of the interaction Hamiltonian (\ref{hamint}) follows the fact that the system (an atom or a cavity mode) can absorb at a given moment  at most one photon. Lack of an interaction between the system and the output field means that the photons emitted by the system leaves the interaction region and can not be reabsorbed. We describe here briefly the {\it a priori} and the {\it a posteriori} evolution of a cavity mode coupled to a propagating one-dimensional Bose field in a superposition of two coherent states. Thus, we have
\begin{equation}
H_{\mathcal{S}}=\omega_{0}a^{\dagger}a,
\end{equation}
\begin{equation}
L=\sqrt{\Gamma} a,
\end{equation}  
where $a$ stands for the
annihilation operator, $\omega_{0}>0$, and $\Gamma$ is the positive coupling constant. We consider here the case when the harmonic oscillator is initially in the coherent state
\begin{equation}
a|u\rangle=u|u\rangle.
\end{equation}
Then, the solution to the set of the master equations can be written in the form 
\begin{equation}\label{apriori}
\varrho_{t}=|c_{\alpha}|^2|f_{t}\rangle\langle f_{t}|+c_{\alpha}c_{\beta}^{\ast}\frac{\langle \beta|\alpha\rangle}{\langle g_{t}| f_{t}\rangle}
|f_{t}\rangle\langle g_{t}|+
c_{\alpha}^{\ast}c_{\beta}\frac{\langle \alpha|\beta\rangle}{\langle f_{t}| g_{t}\rangle}
|g_{t}\rangle\langle f_{t}|+|c_{\beta}|^2|g_{t}\rangle\langle g_{t}|,
\end{equation}
where
\begin{equation}
\langle g_{t}| f_{t}\rangle=\exp\left\{-\frac{1}{2}\left(|g_{t}|^2+|f_{t}|^2-2g_{t}^{\ast}f_{t}\right)\right\},
\end{equation}
and $|f_{t}\rangle$, $|g_{t}\rangle$ are coherent states of the harmonic oscillator with the amplitudes satisfying the equations
\begin{equation}
\dot{f}_{t}=-\left(\mathrm{i}\omega_{0} +{\Gamma \over 2}\right)f_{t} 
-\sqrt {\Gamma }\alpha_{t},
\end{equation}
\begin{equation}
\dot{g}_{t}=-\left(\mathrm{i}\omega_{0} +{\Gamma \over 2}\right)g_{t} 
-\sqrt {\Gamma }\beta_{t},
\end{equation}
where one can easily recognize the damping and driving terms.
Hence, we obtain 
\begin{equation}\label{f}
f_{t}=e^{-\left(\mathrm{i}\omega_{0} +{\Gamma \over 2}\right)t} 
\left( u -\sqrt {\Gamma }\int_{0}^{t}
e^{\left(\mathrm{i}\omega_{0} +{\Gamma \over 2}\right)s}
\alpha_{s} ds \right)
\end{equation}
\begin{equation}\label{g}
g_{t}=e^{-\left(\mathrm{i}\omega_{0} +{\Gamma \over 2}\right)t} 
\left( u -\sqrt {\Gamma }\int_{0}^{t}
 e^{\left(\mathrm{i}\omega_{0} +{\Gamma \over 2}\right)s}
\beta_{s} ds \right).
\end{equation}
The solution (\ref{apriori}) one can check simply by inserting it into Eqs. (\ref{master1})-(\ref{master3}).  
The conditional state of the system can be written as
\begin{equation}\label{posteriori}
\tilde{\rho}_{t}=|c_{\alpha}|^2G^{\alpha\alpha}_{t}|f_{t}\rangle\langle f_{t}|+c_{\alpha}c_{\beta}^{\ast}\frac{\langle \beta|\alpha\rangle}{\langle g_{t}| f_{t}\rangle}
G^{\alpha\beta}_{t}|f_{t}\rangle\langle g_{t}|+
c_{\alpha}^{\ast}c_{\beta}\frac{\langle \alpha|\beta\rangle}{\langle f_{t}| g_{t}\rangle}
G^{\beta\alpha}_{t}|g_{t}\rangle\langle f_{t}|+|c_{\beta}|^2G^{\beta\beta}_{t}|g_{t}\rangle\langle g_{t}|,
\end{equation}
where the conditional amplitudes $f_{t}$ and $g_{t}$ coincide with the {\it a priori} ones given by (\ref{f}) and (\ref{g}), and the stochastic coefficients $G^{\alpha\alpha}_{t}$, $G^{\alpha\beta}_{t}$, $G^{\beta\alpha}_{t}$, and $G^{\beta\beta}_{t}$ for the counting stochastic process satisfy the equations
\begin{equation}
dG^{\alpha\alpha}_{t}=\left(\nu_{t}^{\alpha\alpha}-G^{\alpha\alpha}_{t}\nu_{t}\right)\left(dn_{t}-\nu_{t}dt\right),
\end{equation}
\begin{equation}
dG^{\alpha\beta}_{t}=\left(\frac{\nu_{t}^{\alpha\beta}}{\langle \beta|\alpha\rangle }-G^{\alpha\beta}_{t}\nu_{t}\right)\left(dn_{t}-\nu_{t}dt\right),
\end{equation}
\begin{equation}
dG^{\beta\alpha}_{t}=\left(\frac{\nu_{t}^{\beta\alpha}}{\langle \alpha|\beta\rangle }-G^{\beta\alpha}_{t}\nu_{t}\right)\left(dn_{t}-\nu_{t}dt\right),
\end{equation}
\begin{equation}
dG^{\beta\beta}_{t}=\left(\nu_{t}^{\beta\beta}-G^{\beta\beta}_{t}\nu_{t}\right)\left(dn_{t}-\nu_{t}dt\right),
\end{equation}
where the intensities have the form
\begin{equation}
\nu_{t}^{\alpha\alpha}=\left|\sqrt{\Gamma}f_{t}+\alpha_{t}\right|^2G^{\alpha\alpha}_{t},
\end{equation}
\begin{equation}
\nu_{t}^{\alpha\beta}=\left(\sqrt{\Gamma}f_{t}+\alpha_{t}\right)\left(\sqrt{\Gamma}g_{t}^{\ast}+\beta_{t}^{\ast}\right)G^{\alpha\beta}_{t}\langle\beta|\alpha \rangle,
\end{equation}
\begin{equation}
\nu_{t}^{\beta\alpha}=\left(\sqrt{\Gamma}f_{t}^{\ast}+\alpha_{t}^{\ast}\right)\left(\sqrt{\Gamma}g_{t}+\beta_{t}\right)G^{\beta\alpha}_{t}\langle\alpha|\beta \rangle,
\end{equation}
\begin{equation}
\nu_{t}^{\beta\beta}=\left|\sqrt{\Gamma}g_{t}+\beta_{t}\right|^2G^{\beta\beta}_{t}.
\end{equation}
For the homodyne observation, we get
\begin{equation}
dG^{\alpha\alpha}_{t}=\left(\mu_{t}^{\alpha\alpha}-G^{\alpha\alpha}_{t}\mu_{t}\right)\left(dq_{t}-\mu_{t}dt\right),
\end{equation}
\begin{equation}
dG^{\alpha\beta}_{t}=\left(\frac{\mu_{t}^{\alpha\beta}}{\langle \beta|\alpha\rangle }-G^{\alpha\beta}_{t}\mu_{t}\right)\left(dq_{t}-\mu_{t}dt\right),
\end{equation}
\begin{equation}
dG^{\beta\alpha}_{t}=\left(\frac{\mu_{t}^{\beta\alpha}}{\langle \alpha|\beta\rangle }-G^{\beta\alpha}_{t}\mu_{t}\right)\left(dq_{t}-\mu_{t}dt\right),
\end{equation}
\begin{equation}
dG^{\beta\beta}_{t}=\left(\mu_{t}^{\beta\beta}-G^{\beta\beta}_{t}\mu_{t}\right)\left(dq_{t}-\mu_{t}dt\right),
\end{equation}
where
\begin{equation}
\mu_{t}^{\alpha\alpha}=\left(\sqrt{\Gamma}\left(f_{t}+f_{t}^{\ast}\right)+\alpha_{t}+\alpha_{t}^{\ast}\right)G^{\alpha\alpha}_{t},
\end{equation}
\begin{equation}
\mu_{t}^{\alpha\beta}=\left(\sqrt{\Gamma}\left(f_{t}+g_{t}^{\ast}\right)+\alpha_{t}+\beta_{t}^{\ast}\right)G^{\alpha\beta}_{t}\langle\beta|\alpha \rangle,
\end{equation}
\begin{equation}
\mu_{t}^{\beta\alpha}=\left(\sqrt{\Gamma}\left(f_{t}^{\ast}+g_{t}\right)+\alpha_{t}^{\ast}+\beta_{t}\right)G^{\beta\alpha}_{t}\langle\alpha|\beta \rangle,
\end{equation}
\begin{equation}
\mu_{t}^{\beta\beta}=\left(\sqrt{\Gamma}\left(g_{t}+g_{t}^{\ast}\right)+\beta_{t}+\beta_{t}^{\ast}\right)G^{\beta\beta}_{t}.
\end{equation}
One can prove (\ref{posteriori}) by inserting the conditional operators
$\tilde{\rho}_{t}^{\alpha\alpha}$, $\tilde{\rho}_{t}^{\alpha\beta}$, $\tilde{\rho}_{t}^{\beta\alpha}$, and $\tilde{\rho}_{t}^{\beta\beta}$ of the proposed forms into the relevant filtering equations. One can check that this leads to the given differential equations for the amplitudes $f_{t}$ and $g_{t}$, and the coefficients $G^{\alpha\alpha}_{t}$, $G^{\alpha\beta}_{t}$, $G^{\beta\alpha}_{t}$, and $G^{\beta\beta}_{t}$. For the environment taken in a coherent state our formulas reduces to the known results (see, for instance, \cite{BB91,Car02}).

\section{Conclusion} 

We derived the stochastic equation describing the conditional evolution of an open quantum system interacting with the Bose field prepared in a superposition of coherent states. We consider two schemes of measurement of the output field: photon counting and homodyne detection. Instead of methods based on the quantum stochastic calculus and the cascades system model \cite{GJN12b}, we used the collision model with the environment given by an infinite chain of qubits. We assumed that the bath qubits do not interact between themselves and they are initially prepared in the entangled state being a discrete analogue of a superposition of continuous-mode coherent states. The initial state of the compound system was factorisable. Because of the temporal correlations present in the environment, the evolution of open quantum system becomes non-Markovian. We started from the discrete in time description of the problem and obtained in the continuous time limit differential filtering equations consistent with the results published in \cite{GJNC12a, GJN13}. We would like to stress that the presented method is more straight and intuitive than the methods described in \cite{GJNC12a, GJN13}. It not only allows to derive the equations describing the conditional evolution of the system but also enables to find the general structure of quantum trajectories.

\section*{Acknowledgements}

This paper was partially supported by the National Science Center project 2015/17/B/ST2/02026.

\end{document}